\documentclass[11pt]{article}
\usepackage{array}
\usepackage{graphicx}
\usepackage{cite}
\usepackage{textpos}
\usepackage{bbold}
\usepackage{adjustbox}

\title{Free Will and Falling Cats}
\author {Frank Wilczek  \\
\small\it Center for Theoretical Physics, MIT, Cambridge, MA 02139 USA; \\
\small\it T. D. Lee Institute and Wilczek Quantum Center, \\
\small\it Shanghai Jiao Tong University, Shanghai, China;\\
\small\it Arizona State University, Tempe, AZ, USA; \\
\small\it Stockholm University, Stockholm, Sweden }

\begin{document}

\maketitle

\begin{textblock*}{5cm}(11cm,-8.2cm)
\fbox{\footnotesize MIT-CTP/5718}
\end{textblock*}

\begin{abstract}
If we consider a cat  to be an isolated  mechanical system governed by T-invariant mechanics, then its ability to land on its feet  after being released from rest is incomprehensible.  It is more appropriate to treat the cat as a creature that can change its shape in order to accomplish a purpose.  Within that framework we can construct a useful and informative of the observed motion.   One can learn from this example.  
\end{abstract}

\medskip

\section{The ``Bad News for Cats'' Theorem}

The ability of cats to recover from a sudden fall from a height to land on their feet is legendary.  The related ability of trained human divers and gymnasts to stick an intended landing is at least as impressive, and much more reliably documented.  These feats have attracted the interest of Maxwell \cite{Maxwell} and of many other physicists -- not to mention athletes and trainers -- down to recent times \cite{recent_cats} \cite{recent_divers}.

In 1904 the renowned mathematician Paul Painlev\'e contributed a short, rather cryptic note \cite{painleve} announcing the following result:
\begin{quote}
... let S be a conservative system in which each element remains
identical to itself, so that the state of the system at an instant t is completely
defined by the position and velocity of each element. Given that this system S is
abandoned without velocity in a vacuum above the ground, is it possible for it to
return to its initial configuration, oriented DIFFERENTLY in space, at an instant
t?

This is a question that has already been discussed before at the Academy, a few
years ago (the problem of the cat that lands on its paws). When we don't subjugate
a subject the system S to being conservative, the answer is affirmative and has been
supported by numerous examples. 

On the contrary, when S is conservative, the answer to the question asked is
negative: if S returns to its initial configuration, it is surely oriented in the same
way in space. This is a remarkable fact ...
\end{quote}

As the subsequent text makes clear, by ``conservative system'' Painlev\'e means a system point of masses  that interact through central potentials according to the laws of Newtonian mechanics.   
Later reconstructions of the proof \cite{Geru} \cite{Sachs} are rather complicated.   Here, working at a more abstract level, we give a simple conceptual proof of a more general result:

\bigskip

{\it Theorem}: (a) If a closed system governed by time-reversal invariant dynamics assumes time-reversal even states $S(t_0)$, $S(t_0 + \tau)$ at two times $t_0$, $t_0 + \tau$ separated by an interval $\tau$, then its motion is periodic with period $2\tau$.  (b) If, furthermore, those states are related to one another by a transformation $R$ that is a symmetry of the dynamical evolution, then $R^2$ acting on $S(t_0)$ (and on $S(t_0+ \tau)$) is trivial.  

{\it Proof}: (a): By applying time reflection around $t_0$ we have $S(t_0 + t ) = S(t_0 - t)$, and by applying time reflection around $t_0 + \tau$ we have $S(t_0 -t ) = S(t_0 + \tau + (-\tau -t)) = S(t_0 + \tau - (-\tau - t)) = S(t_0 + t + 2\tau )$.   Thus $S(t_0 + t) = S(t_0 + t + 2\tau$).  

(b): If we denote the operation that evolves states through time $t$ by $U(t)$, so $U(t) S(t_i) \equiv S(t_i + t)$, then the mathematical formulation of the statement that $R$ is a symmetry of the dynamics is $RU(t) = U(t)R$.  So if  $S(t_0 + \tau) = RS(t_0)$ and $R$ is a symmetry of the dynamics, then $S(t_0) = S(t_0 + 2 \tau) = U(\tau) S(t_0 + \tau) = U(\tau) R S(t_0) = RU(\tau)S(t_0) = R^2 S(t_0)$.  

\bigskip

Probably the most interesting application of  part (b) concerns symmetry under rotation (including, for parity-invariant forces, improper rotation).   In those cases, $R^2 S = S$ is a severe constraint on $RS$, basically limiting it to $\pi$ rotations or (assuming parity symmetry) reflections and inversions.  

Time reversal symmetry is essential to the theorem, as is shown in the Appendix.  

Painlev\'e's interest in this problem was stimulated by the falling cat problem, and this theorem has been called the ``Bad News for Cats Theorem" \cite{roberts}\cite{earman}.   Taken at face value, of course, the theorem makes cats' achievements seem paradoxical, since it contradicts their observed ability to re-orient.   

\section{Why Cats are OK} 

I suspect that everyone involved in this discussion has had tongue firmly planted in cheek, but at the risk of belaboring the obvious I want to highlight why this theorem is not relevant to real biological cats.  While simple, the reason touches on profound issues, as signaled by the title of this note. 

The point is that the framework assumed in the theorem, even in its generalized form, is not appropriate for the description of the behavior of biological cats, divers, or gymnasts.  Specifically, none of those are closed systems, nor is their state time-reversal invariant even when they appear, macroscopically, to be at rest.  They can readily and selectively consume stored energy, notably by converting ATP into ADP, empowering mechanical motion accompanied by radiation of heat.   Indeed, while they are living, they never cease doing that.  Also, they can use stored energy to process and transmit information, and to move in response to its flow.  Thus, the assumptions of the theorem do not apply to biological actors.

\section{Engaging With Will}

Fortunately, a radically different approach offers useful  insight into this class of problems.  That is, we assume that the cat, diver, or gymnast has significant abilities to change its shape, by ``will'', in response to external cues and internal plans.  Within that framework many fascinating questions about what degrees of freedom can be called into play, how quickly they respond, how the systems can learn by practice and instruction, and so forth, in order to attain specific goals, can be discussed usefully.   This is the approach adopted universally, though usually implicitly, by cats, biologists, athletes, and trainers.  On the other hand it is deprecated, in effect if not by intention, by physicists and philosophers who deny the reality of will.  

The utility of the concept of will, in this context, does not mean that anything goes.   The ``Astonishing Hypothesis'' \cite{crick} that mind (of which will is an aspect) is an emergent property of matter is not here challenged. Instead, a broad-minded approach, that takes into account both the legitimacy of will and constraints that follow from the laws of physics, is most fruitful.  For example, conservation of angular momentum holds to a good approximation for cats and divers in flight, since it is difficult to radiate away angular momentum.  The kinematics of deformable bodies under that constraint is already quite rich \cite{recent_cats} \cite{recent_divers}, and naturally brings in ideas from non-abelian gauge theory \cite{fwDeformable}.   It helps us to track precisely how changes of shape induce changes in orientation.  

Engineers rarely speak of ``will'' or ``purpose'',  but they speak frequently of ``control'' and ``utility''  \cite{feedback} \cite{control}, which are parallel concepts.    Recently it has become increasingly common also to speak of ``intelligent'' engineered systems.  These quoted concepts are extra-physical ideas, yet they are central to the process of designing complex material systems and anticipating their behavior.   And for those who observe the finished products, they are essential tools for understanding how they got to be the way they are and for making sense of their structure.  

Complementarity, as discussed in \cite{fwFundamentals}, is the insight that very different and even superficially contradictory ways of describing the same system can provide ways to answer different questions about it.   These uses of extra-physical concepts to address questions about the physical behavior of material systems nicely illustrate the utility of complementarity in action.  

\bigskip

The preceding considerations about will and movement to achieve re-orientation apply to motion more generally.  
In everyday life most of our experience of motion is associated with the movement of our own bodies.   We plan motions and will make them happen, for example in rising from sleep, walking, bringing food to our mouths, and a host of other activities that are essential to life.   This experienced mechanics is superficially -- and also deeply -- different from classical Newtonian mechanics.  We can change our arm's velocity, for example, in apparent violation of the conservation of energy, or accelerate it without applying external forces (external, that is, to our bodies). 

A big part of the reason why classical Newtonian mechanics was difficult to discover, and a big part of why it is such a towering intellectual achievement, is that it required a kind of cognitive dissonance.  Specifically, it required -- and still requires --  people to reconcile two very different, quasi-contradictory descriptions of motion, both successful and convincing within their domains of application.  There is the ``common sense'' understanding of movement, in which the concepts of will and purpose are central, that applies to experienced bodily activity; and there is classical mechanics proper, austere and rigorously mathematical, from which those concepts are excluded.   

A few centuries ago, when classical mechanics first took shape, the difference between those descriptions was widely believed to be fundamental.  The first description applied when souls or other animating principles were active, while the other applied while those animating principles were dormant.  Today, science has understood that matter, embodying a system of known, (relatively) simple, mathematically formulated laws, can support the sort of complex emergent behavior we find, for example, in molecular biochemistry, neurobiology and synthetic intelligence.  It has come to seem much more reasonable, therefore, to regard those two descriptions as complementary.  They are alternative descriptions of the same reality, tailored to address different questions.  We can use one or the other, or a hybrid.  The conceptually enriched reality that results makes room for will and purposive motion without bringing in souls or animating principles.

\section{Engaging With Free Will}

I feel that nothing that I have said so far should be controversial.  Nevertheless, it prepares us to address some notoriously contentious issues that have a similar logical structure. 

\subsection{Making Choices}

Another common experience in human life is that of making deliberate choices.  We imagine the consequences of different possible actions, and select which to carry out.   This process can be discussed from many perspectives, that use different concepts and offer different insights.  For present purposes, some brief indications will suffice:
\begin{itemize}
\item Psychology: personality, drives
\item Economics: utility, game theory
\item Anthropology: culture
\end{itemize}

On the other hand one might be tempted, based on a crude interpretation of the Astonishing Hypothesis, to deny the reality of choice altogether.   If mind emerges from matter, and the behavior of matter is determined by a fixed set of deterministic equations then, according to this interpretation,  the whole notion of making choices is illusory.  What is going to happen, happens, full stop.

And yet, to paraphrase Galileo, we choose.  Thousands of books and papers in the above-mentioned disciplines, not to mention history and fiction, discuss the human world in terms of people making choices.  People can have rational discussions using those ideas, at various levels of sophistication, and they can learn to use them better.   

It is hard not to see the tension between the pair (choice)-(physics in general) as a grander version of the tension (will)-(classical mechanics) we saw in our discussion of falling cats.   Whether or not it is possible ``in principle'' to anticipate or understand the behavior of people from the quantum wave function of their constituent quarks, gluons, electrons, and photons, that is rarely the best way to do so.   We can have richer and more informative discussions, and make more and better predictions, if we bring in extra-physical concepts.  

There are many reasons why it is useful to bring in such concepts. For while the fundamental equations of physics, as presently understood, evolve the present state of a closed physical system into a unique state at any future time, that fact has very limited relevance to the question of determinism in human behavior, because:
\begin{itemize}
\item People are very far from being closed systems.  They eat, breathe, absorb and radiate heat, get rained on, and so forth, and -- most important -- sense their environment and process and act upon that sensory information.  
\item Calculating the evolution requires near-complete and accurate knowledge of the initial state, which is not available even for much smaller systems than human bodies.  
\item Even given perfect knowledge of the initial state, the required calculations are absurdly impractical.
\end{itemize}
Thus, it becomes essential to discover and develop concepts that correspond to more robust observables and lead to simpler calculations.

\bigskip

The limitations flowing from the foregoing constraints are so severe as to make the project of calculating human behavior from the fundamental equations preposterous, in practice.   Still, it can be entertaining to ask: What about in theory?  

The first issue can be postponed by taking in larger parts of the universe than individual human beings, so as to achieve effective closure.   But that stratagem dodges rather than resolves the issue of experienced choice -- which is, of course, experienced by individuals.  It also aggravates the other two issues. Moreover, it is unclear that effective closure can ever be achieved in our universe, since a) even small distant events can ramify into big consequences (butterfly effect \cite{butterfly_effect}), and b) according to physical cosmology, at every moment previously unobservable parts of the universe just begin to extend their influence to us (expanding horizon).   

The second issue is at least aggravated, and may be fatally compromised, by quantum theory. In that context, deterministic evolution holds for the wave-function (of a closed system).  But, as articulated most clearly in the many-worlds interpretation of quantum theory, the relationship between the wave-function and experienced reality is not deterministic.  Moreover, since measurements inevitably change the wave-function it is not at all clear that the ideal of complete knowledge of the wave-function corresponds to an operational, empirically based concept. 

The third issue too might pass, upon closer investigation, from severely impractical to strictly impossible.  Computation has a physical basis \cite{feynman} \cite{lloyd}, and the demands of absurdly complex computations might well exceed the resources of the physical universe.   Also, the kind of self-referential computations that figure in paradoxes of the kind ``I compute what I will do, and then will myself to do the opposite'', bear a family resemblance to the paradoxes that underly the famous no-go theorems of G\"odel (incompleteness of deductive systems) and Turing (halting problem), and are likely to be subject to similar limitations.  

\bigskip

For me, the entertainment value of sightseeing in these murky depths stales rapidly, so I will close this section with a related but more nearly grounded speculative comment.

It is possible to imagine that a more advanced neurobiology, powered by better understanding of how the organs of mind function and interact at a molecular level, will contribute incisive new concepts to descriptive psychology.  This, combined with advanced imaging techniques, will usher in new levels of self-awareness.  Specifically, it is not preposterous to imagine a future when people will be able to visualize and meaningfully interpret flows of information within their own or other people's brains in real time, within an immersive, augmented reality.  To do justice to that expanded reality, people will need to develop more refined and possibly radically different concepts around ``choice''.

\subsection{Influencing Choices}

The idea that people can make choices based on considering alternative plans and weighing their merits is basic to morality and law.  Concepts such as guilt, sin, social acceptability, and legality are meant to judge or influence people's choices.   In all these cases, there is an important distinction between {\it acts\/} and {\it choices}.  

For the sake of concreteness, I will focus here on law.  In a world where choice was illusory, the law would only refer to acts, and its only purpose would be retribution.  Issues of intent would not -- indeed, could not -- arise, nor defense by reason of insanity.   Fortunately, we do not live in that world.  (At least, I've argued that we don't.) The law as it actually operates in civilized society is instead largely designed to avoid the occurrence of illegal acts by making them, through the threat of punishment, less attractive choices.  In cases where there is no intent, the law did not influence any choice at all.  Therefore its deterrent purpose was not disrespected, but simply inoperative, and punishment loses some of its point.

\bigskip

The process of making choices is often phrased as ``free will''.  Indeed, the current Wikipedia entry for ``Free Will'' begins
\begin{quote}
Free will is the capacity or ability to choose between different possible courses of action \cite{wikiref}.
\end{quote}
In so far as this is a valid definition, it is not clear that ``free will'' has any advantage over the simpler ``choice''.  The historical origins of ``free will'' are in theology.  In that context its connotations might be welcome.  But in other contexts they are excess baggage.  For the reason that follows, the expression ``free will'' is probably best avoided.

Unfortunately, the phrase {\it free\/} will invites, through its linguistic connotations, extrapolations from its proper use in descriptive psychology and law into dubious speculations or even claims that its utility points to serious oversights in physics and physical cosmology -- specifically, the failure of these subjects to incorporate the direct dynamical influence of choices emanating from minds.  Let me conclude by re-iterating a basic observation \cite{fwFundamentals} that heavily constrains such speculations and claims.   In modern physics we have learned to how to make very precise calculations of many very delicate effects, and to compare them with very precise measurements.  For one example, the form and magnitude of the magnetic field associated to electrons (technically, their magnetic moment) is both theoretically predicted and, independently, measured to accuracy within fractions of a part per billion.   For another, atomic clocks of several different kinds are predicted and experimentally measured to keep time consistently with one another to accuracies that correspond to a fraction of a second per billion years.  In the course of making such accurate measurements, experimenters must pay careful attention to, and account for, many factors, including mechanical alignment and stability of their apparatus, isolation through high vacuum and ultra-low temperature within the experiment's core, electrical isolation of sensors and amplifiers, screening of stray magnetic and electromagnetic fields, shielding from or vetoing on cosmic ray events, and so forth.  But amidst all this care and delicacy it has never proved necessary to take any precautions against the possible influence of what people nearby the experiments -- or, for that matter, anywhere -- are thinking.   Demonstration of effects of that kind, however minute, would be remarkable.  Based on the existing evidence, we can be confident that they are subtle and elusive, if they exist at all.

\bigskip

\bigskip

%{\it Acknowledgement}: The first section partly adapts material from \cite{fw_nobel}.   The second section adapts material from a forthcoming paper with X. Peng, J. Dai, and A. Niemi %%\cite{pdnw}.  This work is supported by the U.S. Department of Energy under grant Contract  Number DE-SC0012567, by the European 
%Research Council under grant 742104, and by the Swedish Research Council under Contract No. 335-2014-7424.

{\it Acknowledgement}: I thank Juliana Baena for introducing me to the literature around Painlev\'e's paper and for preparing a translation (quoted in part above).  Also I thank her and Antti Niemi for discussions about the physics around it.  I thank Brian Greene and Maulik Parikh for discussions about the natural philosophy of will and choice that prompted me to write down some of my thoughts on those matters.  This work is supported by the U.S. Department of Energy under grant Contract  Number DE-SC0012567 and by the Swedish Research Council under Contract No. 335-2014-7424.

\pagebreak

{\it Appendix: Minimal Mechanical Examples; Necessity of T}

\bigskip

We can illustrate the application of our theorem, and see that T symmetry is essential, by considering a particle subject to the simple Lagrangian 
\begin{equation}
L ~=~ \frac{1}{2} ({\dot r}^2 + r^2 {\dot \phi}^2) \, -  \, V(r) \, + \, \alpha f(r) \dot \phi
\end{equation}
in polar coordinates. This Lagrangian describes planar motion for a unit charge, unit mass particle in the presence of a transverse, axially symmetric magnetic field $B_z(r) = \alpha \frac{1}{r} \frac{df}{dr}$

For the cyclic variable $\phi$ we have the equation of motion $\frac{d}{dt}
\frac{\partial L}{\partial \dot \phi} = 0 $, so 
\begin{equation}
p_\phi ~=~ \frac{\partial L}{\partial \dot \phi } = r^2 \dot \phi + \alpha f(r) 
\end{equation}
is a constant of the motion.  $T$-invariant states are defined by varnishing velocity, including $\dot \phi = 0$, so 
\begin{equation}
p_\phi ~\rightarrow~ \alpha f(r_0)
\end{equation}
and 
\begin{equation}
\dot \phi = \frac{\alpha(f(r_0) - f(r))}{r^2}
\end{equation}

T invariance of the dynamics implies $\alpha = 0$, and thus $\dot \phi = 0$. If the particle does not pass through the origin $r=0$, then $\phi$ is uniquely defined and constant, so the orientation of our particle cannot change, as is consistent with the implication b. of our theorem.   If the particle passes through the origin with non-zero velocity then to maintain continuity of the velocity we must allow $\phi$ to increment by $\pi$.  Following that the next turning point might occur at the same value of $r$ and be related to the original one by a $\pi$ rotation.  This is consistent with the theorem, and illustrates that we must allow for the possibility of a non-trivial $R$. 

Essentially this same example can be realized within the context of Painlev\'e's original framework of particles and central forces, by having two very heavy particles acting equally on a third light particle equally distant from both.  This model demonstrates, in particular, that $R \neq 1$ can occur without collisions.  That possibility contradicts \cite{painleve} \cite{Geru} \cite{Sachs}.  

With $\alpha \neq 0$ things are different. We can derive the equation of motion for $r$ straightforwardly, or more elegantly using the Routhian 
\begin{equation}
R ~=~ - \frac{1}{2} {\dot r}^2 + \frac{1}{2} \frac{(p_\phi- \alpha  f(r))^2}{ r^2 }+ V(r)
\end{equation}
in the form
\begin{eqnarray}
\frac{d}{dt}\frac{\partial R}{\partial \dot r} ~&=&~ \frac{\partial R}{\partial r} \nonumber \\
\ddot r  ~&=&~  - \frac{d}{dr} (V(r) + \frac{1}{2} \frac{(p_\phi- \alpha  f(r))^2}{ r^2 } )
\end{eqnarray}
Using the conserved effective energy
\begin{equation}
{\cal E} ~=~ \frac{1}{2} {\dot r}^2 + V(r) + \frac{1}{2} \frac{(p_\phi- \alpha  f(r))^2}{ r^2 }
\end{equation}
we see that resting points, where $\dot r = 0$, generally represent places where the $r$-evolution reverses direction.  If there is a second resting point on the same trajectory, we will have oscillations bounded by the resting points.   Generically, the accumulated rotation
\begin{equation}\label{phase_accumulation}
\Delta \phi ~=~   \oint \frac{\alpha(f(r_0) - f(r))}{r^2} dt
\end{equation}
over a cycle will be neither $0$ nor $\pi$, and we can encounter re-orientations (i.e., in this context, rotations) of any magnitude.  

For example, let us take a harmonic oscillator potential $V(r) = \frac{1}{2}kx^2$ and a constant magnetic field $\alpha f(r) = \frac{1}{2}B_0 r^2$.  If at $r= r_0$ we have $\dot r = \dot \phi = 0$ then 
\begin{eqnarray}
{\cal E} ~&=&~ \frac{1}{2} k r_0^2 \nonumber \\
p_\phi ~&=&~ \frac{B_0}{2} r_0^2
\end{eqnarray}
and 
\begin{equation}
{\dot r}^2 ~=~ k(r_0^2 - r^2) - \frac{B_0^2}{4} \frac{(r_0^2 - r^2)^2}{r^2}
\end{equation}
There is a turning point where $\dot r = 0$ at 
\begin{equation}
r_1 ~=~ \frac{r_0}{\sqrt{1 +\frac{4k}{B_0^2}}}
\end{equation}
(Note that $\dot \phi$ does {\it not\/} vanish when $r = r_1$.)   When $r$ cycles back to $r_0$, the accumulated phase is, using $dt = \pm \frac{dr}{\dot r}$ appropriately in Eqn.\,(\ref{phase_accumulation}), 
\begin{equation}
\Delta \phi ~=~ -2 \frac{B_0}{|B_0|} \int\limits^{r_0}_{r_1} \frac{dr}{r} \frac{1}{\sqrt{1 + \frac{4k}{B_0^2}}}\sqrt{\frac{r_0^2 -r^2}{r^2- r_1^2}}
\end{equation}
Since the integrand is positive, this does not vanish. 

\end{document}